\documentstyle[sprocl]{article}

\input{psfig}

\newcommand \bra[1]{\left< {#1} \,\right\vert}
\newcommand \ket[1]{\left\vert\, {#1} \, \right>}
\newcommand \braket[2]{\hbox{$\left< {#1} \,\vrule\, {#2} \right>$}}
\newcommand{\bea}{\begin{eqnarray}}
\newcommand{\eea}{\end{eqnarray}}
\newcommand{\simgt}{\hbox{ \raise3pt\hbox to 0pt{$>$}\raise-3pt\hbox{$\sim$} }}
\newcommand{\simlt}{\hbox{ \raise3pt\hbox to 0pt{$<$}\raise-3pt\hbox{$\sim$} }}

\begin{document}

\title{Gauge dependence and matching procedure of a nonrelativistic\\
 QCD boundstate formalism
\footnote{Talk given at the International
Symposium on ``Quantum Chromodynamics and Color Confinement''
(Confinement 2000),
Osaka, Japan, March, 2000.}
}

\author{Y. Sumino}

\address{Department of Physics, Tohoku University\\
    Sendai, 980-8578 Japan} 


\maketitle\abstracts{
We investigate gauge dependence of a nonrelativistic boundstate formalism 
used in contemporary calculations. It is known that the effective 
Hamiltonian of the boundstate system depends on the choice of gauge.  We 
obtain the gauge transformation charge of the Hamiltonian, by which gauge 
independence of the mass spectrum and gauge dependences of the boundstate 
wave functions are dictated.  We raise two questions of practical and
physical interest, and provide answers
to them.
}

\section{Introduction}

Recently there has been much progress in our theoretical
understanding of 
nonrelativistic QED and QCD (NRQED and NRQCD) boundstates such as
positronium, $\Upsilon$ and remnant of toponium boundstates
\cite{cl}$^-$\cite{upsilon}.
A notable characteristic in these new developments is that 
the conventional Bethe-Salpeter equation is no longer being used
to calculate the spectrum and wave functions of boundstates. 
Instead, one starts from the non-relativistic Schr\"odinger equation
(of quantum mechanics) with the Coulomb potential.
Then one adds to the nonrelativistic 
Hamiltonian relativistic corrections and radiative corrections as
perturbations to obtain an effective Hamiltonian 
(quantum mechanical operator) valid up to a necessary
order of perturbative expansion.
Effective Hamiltonians used in these new formalisms are known to be
dependent on the choice of gauge.

We report here our recent achievements \cite{hasebesumino}
on investigations of
gauge dependence of a
nonrelativistic boundstate formalism used in contemporary calculations.
Our motivations for the study are:
(I)
In the present frontier calculations of higher order corrections
to physical quantities of boundstates, often the Feynman gauge is
used to calculate typically ultraviolet radiative corrections
whereas the Coulomb gauge is used to calculate corrections
originating typically from infrared regions.
Therefore, it is desirable to clarify gauge dependences
of the formalisms actually used in these calculations.
(II)
We would like to find transformations of boundstate wave functions
when we change the gauge-fixing condition.
We may apply these transformations to study various amplitudes involving
boundstates.
Since a physical amplitude is gauge independent, once
we know how the wave function transforms, we know how
other parts of the amplitude should transform to cancel gauge
dependence as a whole.
This would provide a useful cross check for identifying all 
the contributions that have to be taken into account at a 
given order of perturbative expansion.

A formal argument of gauge dependence/independence of boundstate
formalisms goes as follows.
It is well known that a quark-antiquark boundstate contributes a pole 
to the quark-antiquark 4-point Green function:
\bea
G(p \, qP) = \frac{i}{2\omega_{\nu ,\vec{P}}} \,
\frac{\chi_{\nu,\vec{P}}(p)\overline{\chi}_{\nu,\vec{P}}(q)}
{P^0 - \omega_{\nu ,\vec{P}} + i \epsilon }
\, + (\mbox{regular as }P^0 \to \omega_{\nu ,\vec{P}} ) ,
\label{pole}
\eea
An infinitesimal deformation of the gauge-fixing function
in the QCD Lagrangian,
$F \to F + \delta F$, induces a variation
\bea
\int d^4x \, \delta {\cal L} = \{ i Q_B , \delta {\cal O} \} ,
~~~~~ ~~~~~
\delta {\cal O} \equiv \int d^4x \, {\rm tr} [ \bar{c} \, \delta F ] .
\eea
Any physical amplitude $\braket{f;{\rm out}}{i;{\rm in}}$ which involves
the quark-antiquark boundstate contributions includes the above Green 
function $G(pqP)$ as a part of it.
Since the initial and final states satisfy the physical state conditions
$Q_B \ket{i;{\rm in}} = Q_B \ket{f;{\rm in}} = 0$ and the theory
is BRST invariant, the amplitude is gauge independent.
Hence, the boundstate poles included in the amplitude 
are also gauge independent.

Nevertheless we may raise two intriguing questions:
(I)
Does the Green function $G(pqP)$ 
include any unphysical pole, which does
not contribute to the physical amplitude, close to or degenerate
with one of the physical boundstate poles?
A typical example is the $R_\xi$-gauge for electroweak
interaction where an unphysical pole $(k^2-\xi M_W^2 + i\epsilon )^{-1}$
is included in the gauge boson propagator.
(II)
Is the boundstate wave function a physical observable?
The answer is obviously no since the wave function is gauge dependent
as a consequence of the gauge dependence of the effective
Hamiltonian.
Then, what is the (gauge-independent) 
physical quantity which can be thought as a
counterpart of a boundstate wave function?

We will answer to these questions below.

\section{Effective Hamiltonian and Transformation Charge}

Let us sketch the outline of our argument.
The quantum mechanical Hamiltonian is determined from
perturbative QCD order by order in expansion in $1/c$
(inverse of the speed of light):
\bea
\hat{H} =  \hat{H}_0 + \frac{1}{c} \, \hat{H}_1
+ \frac{1}{c^2} \, \hat{H}_2 + \cdots .
\label{nrhamiltonian}
\eea
Since quark and antiquark inside a heavy quarkonium are
non-relativistic, the expansion in $1/c$ leads to a
reasonable systematic approximation.
Presently the Hamiltonian is known up to ${\cal O}(1/c^2)$.
There exist several different definitions of an
effective Hamiltonian for the NRQCD boundstates beyond leading order.
We introduce an effective Hamiltonian defined naturally
in the context of time-ordered (or ``old-fashioned'') perturbation 
theory of QCD.
Then we obtain a transformation charge $Q$ (quantum mechanical operator)
such that the effective Hamiltonian and the boundstate wave function 
change as
\bea
&&
\delta H_{\rm eff}(P^0) = 
\left[ H_{\rm eff}(P^0) - P^0 \right] \, i Q (P^0)
- i Q^\dagger (P^0) \, \left[ H_{\rm eff}(P^0) - P^0 \right]  ,
\nonumber \\
&&
\delta \varphi = - i \,Q \cdot \varphi
\nonumber
\eea
when the gauge-fixing condition is varied infinitesimally.
Written in terms of the BRST charge and the field operators in
the QCD Lagrangian, $Q$ is given by
\bea
&&
\bra{\vec{p},-\vec{p},\lambda,\bar{\lambda}}
Q (P^0) 
\ket{\vec{q},-\vec{q},\lambda',\bar{\lambda}'}
\nonumber \\ &&
= \int \frac{d^3 \vec{q}\, '}{(2\pi)^3} \sum_{\lambda'',\bar{\lambda}''} \,
\bra{\vec{p},-\vec{p},\lambda,\bar{\lambda}} 
Q_B \, \frac{1}{P^0-H+i\epsilon}
\,  \delta O  \,
\frac{1}{P^0-H+i\epsilon}
\ket{\vec{q}\, ',-\vec{q}\, ',\lambda'',\bar{\lambda}''} 
\nonumber \\ &&
~~~~~~~~~~~~~
~~~~~~~~~~~~~
\times
{\cal G}^{-1} 
(\vec{q}\, ',\vec{q};\lambda'',\bar{\lambda}'',\lambda',\bar{\lambda}';P^0).
\nonumber
\eea
Gauge independence of the spectrum can be shown using the
transformation.
Since $Q$ has no pole $\sim (E-M_\nu + i\epsilon)^{-1}$,
there is no unphysical state which
contributes a pole to the Green function ${\cal G}$ that is
degenerate with or close to one of
the poles of the physical boundstates of our interest.
Stating more explicitly, 
there is no unphysical boundstate with a binding energy
$\sim \alpha_S^2 m$.

\section{Application}

It is known that
the top quark momentum distribution in
the $t\bar{t}$ threshold region at leading order 
is proportional to the absolute square of
the wave functions of (would-be) toponium boundstates in momentum 
space \cite{sumino}.
The momentum distribution will be measured at future collider
experiments, so we will be able to probe the boundstate wave functions.
As we have shown in Ref.\cite{hasebesumino}, wave functions of boundstates 
are gauge dependent beyond leading order.
We verified 
that this gauge dependence is cancelled by that of 
the final-state interaction diagrams at ${\cal O}(1/c)$ when
calculating the top quark momentum distribution.
In other words, a boundstate wave function mixes with the final-state
interaction diagrams by gauge transformation.

Physically $b$ quark emitted in top decays carries the color
charge and inevitably interacts with the gluons which were also
responsible for the boundstate formation.
Since the top quark momentum is reconstructed from the momenta of
its offsprings, it is natural that the QCD interaction of
the offsprings mixes via gauge transformation in the determination
of top momentum distribution.

At the same time, this shows that the present calculations\cite{momdist}
of the top momentum 
distribution at ${\cal O}(1/c^2)$ 
are gauge dependent, i.e.\ they vary if we transform the gauge infinitesimally
from the Coulomb gauge, since they do not include the final-state interaction
diagrams.
The example at ${\cal O}(1/c)$ suggests how
gauge cancellations should take place in the complete amplitude at
${\cal O}(1/c^2)$ which has not been obtained yet.

\section{Summary}

\begin{itemize}
\item
We used the BRST symmetry to formulate our arguments,
which enabled us to discuss gauge dependence
of the NRQCD boundstate formalism 
rigorously.
\item
Presently, there exist several different definitions of an
effective Hamiltonian beyond leading order.
We introduced an effective Hamiltonian defined naturally
in the context of time-ordered (or ``old-fashioned'') perturbation 
theory of QED/QCD.
Then we obtained a transformation charge $Q$ (quantum mechanical operator)
when the gauge-fixing condition is varied infinitesimally.
Also, gauge independence of the spectrum is shown using the
transformation.

\item
For illustration: (1)we calculated the transformation charge $Q$ at
next-to-leading order;
(2)we demonstrated gauge cancellations among diagrams 
by examining an infinitesimal gauge transformation of
the amplitude for a $q\bar{q}$ boundstate decaying into 
$q'\bar{q}''W^+ W^-$.
From the latter example, one can deduce that the present calculations of
the top momentum distribution in the $t\bar{t}$ threshold region
at next-to-next-to-leading order
are gauge dependent.

\end{itemize}

\end{document}